\documentclass[12pt,thmsa]{article}
\usepackage{sw20aip}

\input{tcilatex}
\begin{document}

\title{How photon detectors remove vacuum fluctuations }
\author{Emilio Santos \and Departamento de F\'{i}sica. Universidad de Cantabria.
Santander. Spain}
\maketitle

\begin{abstract}
The main argument against the reality of the electromagnetic quantum vacuum
fluctuations is that they do not activate photon detectors. In order to met
this objection I propose a model of photocounting which, in the simple case
of a light signal with constant intensity, predicts a counting rate
proportional to the intensity, in agreement with the standard quantum result.

\end{abstract}

\section{\protect\smallskip Introduction}

The existence of vacuum fluctuations is a straightforward consequence of
field quantization \cite{milonni}. In addition, quantum vacuum fluctuations
have consequences which have been tested empirically. For instance, the
vacuum fluctuations of the electromagnetic field (or zeropoint field, ZPF)
give rise to the main part of the Lamb shift \cite{lamb} and to the Casimir
effect \cite{cm}. The ZPF was proposed in 1912 by Planck when he wrote the
radiation law in the form 
\begin{equation}
\rho \left( \omega ,T\right) =\frac{\omega ^{2}}{\pi ^{2}c^{3}}\left[ \frac{%
\hbar \omega }{\exp \left( \hbar \omega /k_{B}T\right) -1}+\frac{1}{2}\hbar
\omega \right] ,  \label{planck}
\end{equation}
where the second term represents the ZPF. That the thermal spectrum contains
an $\omega ^{3}$ term has been proved by experiments measuring current
fluctuations in circuits with inductance at low temperature \cite{khc}. Of
course, the ZPF term is ultraviolet divergent so that some cutoff should be
assumed, likely at about the Compton wavelength, where the fluctuations of
the Dirac electron-positron sea become important.

It is believed that the ZPF cannot be interpreted as a \textit{real }random
electromagnetic field because it does not activate photodetectors in the
absence of signals. (There is also a gravitational problem because, if the
quantum vacuum fluctuations are at the origin of the cosmological constant
as is usually assumed, the constant should be many orders of magnitude
larger than the observed value. But we shall not be concerned with
gravitational effects in this paper.) A common explanation of the fact that
the ZPF does not activate photodetectors is to say that the ZPF is not 
\textit{real}, but \textit{virtual}. However replacing a word, real, by
another one, virtual, with a less clear meaning is not a good solution. In
the present article I shall show that the behaviour of photodetectors can be
explained without renouncing to the reality of the ZPF. The proof goes via
constructing an explicit model of detector producing a counting rate
proportional to the intensity of the signal, that is able to subtract
efficiently the ZPF.

\section{Stochastic properties of the zeropoint field}

The vacuum field (i. e. the lasts term of eq.$\left( \ref{planck}\right) )$
contains an \textit{average} energy $\frac{1}{2}\hbar \omega $ per normal
mode of the electromagnetic radiation. According to quantum mechanics it is
impossible by any (controllable) means to \textit{reduce} that energy.
Indeed, any reduction would lead to a violation of the Heinsenberg
(uncertainty) relations. Therefore we should assume that vacuum field energy
cannot be changed by the presence of optical devices like mirrors, lenses,
beam-splitters, etc. In fact, they cannot reduce the vacuum energy, as said
above, but they cannot increase it either by conservation of energy. What is
possible is to change the structure of the normal modes and this is what
happens in the Lamb shift (where normal modes are modified by the presence
of an atom) or the Casimir effect (normal modes are changed by macroscopic
objects).

That the vacuum energy cannot be changed by any controllable method also
follows from thermodynamic considerations. In fact the vacuum field exists
even at zero Kelvin, as shown by eq.$\left( \ref{planck}\right) $ and, by
the second law, no useful energy could be extracted from it. We might imagin
a \textit{Maxwell demon }able to change the energy content of some modes of
the vacuum field. But that change would necessarily be uncontrollable. In
summary, we must assume that no action may change the stochastic properties
of the vacuum field (that is, the joint probability distribution of the
energy in the normal modes).

Now we shall derive some relevant properties of the ZPF in free space, that
is far from any material body. The ZPF is characterized by the electric
field, \textbf{E}(\textbf{r},t), and the magnetic field, \textbf{B}(\textbf{r%
},t). For our purposes the most relevant quantity is the radiation intensity
represented by the Poynting vector, 
\[
\mathbf{S(r,}t)=\frac{1}{\mu _{0}}\mathbf{E}(\mathbf{r},t)\times \mathbf{B}(%
\mathbf{r},t). 
\]
At every point in space the components of the Poynting vector of the ZPF may
be considered three independent stochastic processes. Every process should
be stationary with zero mean by symmetry considerations. The autocorrelation
might be derived from the spectrum of the process, which could be easily
related to the last term of $\left( \ref{planck}\right) .$ Thus the most
relevant properties of the vector stochastic process $\mathbf{S(r,}t)$, at
any point in space $\mathbf{r,}$ may be writen 
\begin{equation}
\left\langle \mathbf{S(r,}t)\right\rangle =0,\left\langle S_{j}\mathbf{(r,}%
t)S_{k}\mathbf{(r,}t^{\prime })\right\rangle =\delta _{jk}F\left( t^{\prime
}-t\right) ,  \label{point}
\end{equation}
where $\delta _{jk}$ is the Kronecker delta and $F\left( t^{\prime
}-t\right) $ the autocorrelation function.

Now we shall study the situation where we have a light signal superimposed
to the ZPF. The signal energy is concentrated within a narrow region in
momentum space. However the ZPF would contain an average energy $\frac{1}{2}%
\hbar \omega $ in every mode because that energy cannot be reduced as
explained above. The important result is that the signal frequency bandwidth
is substantially more narrow than the ZPF bandwidth, the latter covering the
whole spectrum until a cut-off. This leads us to approximate the spectrum of
the ZPF by a white noise, which gives an autocorrelation function in the
form of a Dirac\'{}s delta. Thus the properties of the Poynting vector of
the light beam will be (compare with $\left( \ref{point}\right) )$%
\begin{equation}
\left\langle S_{j}\mathbf{(r,}t)\right\rangle =\delta
_{j3}I_{s},\left\langle S_{j}\mathbf{(r,}t)S_{k}\mathbf{(r,}t^{\prime
})\right\rangle =\delta _{j3}\delta _{k3}I_{s}^{2}+\sigma ^{2}\delta
_{jk}\delta \left( t^{\prime }-t\right) ,  \label{poyntin}
\end{equation}
where $S_{3}$ is the component of the Poynting vector in the direction of
the beam, $I_{s}$ the signal intensity and $\sigma $ is a constant. This
equation will be the basis of our subsequent study but I point out that the
approximation $\left( \ref{poyntin}\right) $ may be too crude for some
applications. Possible improvements will be considered elsewhere.

\section{Detection model}

Several models of photodetection have been proposed recently resting upon
the idea that there exists a ''detection time'', T, independent of the light
intensity and such that the probability of a count depends on the radiation
(including the ZPF) which enters the detector during the time T\cite{crs}. I
have shown elsewhere\cite{s} that those models are not compatible with
empirical evidence.

Instead of fixing the detection time, T, I shall assume that a count is
produced when the radiation energy accumulated in the detector surpasses
some threshold. This means that once the photocounter is ready to detect
(which will happen some ''dead time'' after a count is produced, but we will
neglect the dead time here), the detector begins to accumulate the radiation
energy entering in it. If I(t) is the total intensity (we are using the word
intensity for ''component of the Pointing vector in the direction of the
beam'') entering the detector at time t, the accumulated energy at time T
will be 
\begin{equation}
E(T)=A\int_{0}^{T}I(t)dt,  \label{energy}
\end{equation}
where A is the entrance area of the detector (in the following we shall put
A = 1 for the sake of simplicity).

The essential assumption of our model is that \textit{a detection event is
produced at a time T, after the previous count, when T is such that } 
\begin{equation}
E(T)\equiv \int_{0}^{T}I(t)dt=E_{m},  \label{model}
\end{equation}
\textit{where I(t) is the radiation intensity entering the detector and E}$_{%
\mathit{m}}$\textit{\ is a parameter characteristic of the detector. }

The use of eq.$\left( \ref{model}\right) $ may be cumbersome due to the
fluctuations of the ZPF and the signal. Indeed constructing a detailed
detection model on the basis of that equation would require using the theory
of ''first passage time'' for the stochastic process $I(t),$ which has a
finite, nonzero, correlation time. However the problem is dramatically
simplified if we assume that $I(t)$ is a white noise (having a null
correlation time) superimposed to a deterministic signal with constant
intensity $I_{s}$, as in eq.$\left( \ref{poyntin}\right) $, so that the
stochastic process $E(T)$ (see $\left( \ref{model}\right) )$ is a Wiener
(Brownian motion) process.

The calculation of the first-passage time is now easy. We shall begin
solving the diffusion equation 
\begin{equation}
\frac{\partial \rho }{\partial t}=\frac{1}{2}\sigma ^{2}\frac{\partial
^{2}\rho }{\partial E^{2}}-I_{s}\frac{\partial \rho }{\partial E}
\label{ode}
\end{equation}
with an absorbing barrier at $E=E_{m}.$ The result is 
\begin{equation}
\rho \left( E,t\right) =\frac{1}{\sigma \sqrt{2\pi t}}\left\{ \exp \left[ -%
\frac{\left( E-I_{s}t\right) ^{2}}{2\sigma ^{2}t}\right] -\exp \left[ \frac{%
2E_{m}I_{s}}{\sigma ^{2}}-\frac{\left( E-I_{s}t-2E_{m}\right) ^{2}}{2\sigma
^{2}t}\right] \right\} .  \label{roet}
\end{equation}
Hence, if we have a detection event at time t\ = 0, the probability that the
next detection event takes place before time t is 
\begin{equation}
P(t)=1-\int_{-\infty }^{E_{m}}\rho \left( E,t\right) dE.  \label{P}
\end{equation}
The integration is straightforward and we get the following distribution of
first-passage times 
\[
P(t)=\frac{1}{2}erfc\left( \frac{E_{m}-I_{s}t}{\sigma \sqrt{2t}}\right) +%
\frac{1}{2}\exp \left( \frac{2E_{m}I_{s}}{\sigma ^{2}}\right) erfc\left( 
\frac{E_{m}+I_{s}t}{\sigma \sqrt{2t}}\right) , 
\]
the corresponding density being 
\[
\frac{dP(t)}{dt}=\frac{E_{m}}{\sigma \sqrt{2\pi t^{3}}}\exp \left[ -\frac{%
\left( E_{m}-I_{s}t\right) ^{2}}{2\sigma ^{2}t}\right] . 
\]

Our aim is calculating the detection rate, which is the inverse of the mean
first passage time, that is 
\begin{equation}
<t>=\int_{0}^{\infty }t\frac{dP(t)}{dt}dt.  \label{mftp}
\end{equation}
The proof that this average gives the inverse of the detection rate is as
follows. We consider that the detector is active during a very large time
interval. Within it we will have a large number of detection events. Let us
assume, for the sake of clarity, that the time intervals between two
detection events form a discrete sequence t$_{1}$, t$_{2}$, ...t$_{j}$,...
If we have N$_{j}$ time intervals of duration t$_{j}$ then the detection
rate will be 
\begin{equation}
R=\frac{\sum N_{j}}{\sum N_{j}t_{j}}=\frac{1}{\sum P_{j}t_{j}}=\frac{1}{<t>},
\label{R}
\end{equation}
where P$_{j}$ is the probability that a time interval between two detection
events has duration t$_{j}$. If we pass to the continuous, we shall replace
the summation by an integral, giving a rate R equal to the inverse of 
\TEXTsymbol{<}t\TEXTsymbol{>}, which completes the proof$.$

The integral in $\left( \ref{mftp}\right) $ may be reduced to standard form%
\cite{grad} with the change of variables x = $E_{m}/\sigma \sqrt{2},$ y = $%
I_{s}/\sigma \sqrt{2},$ u = $\sqrt{t},$ and we obtain 
\[
<t>=\frac{2x}{\sqrt{\pi }}\int_{0}^{\infty }du\exp \left[ -\left( \frac{x}{u}%
-yu\right) ^{2}\right] =\frac{x}{y}. 
\]
Hence eq.$\left( \ref{R}\right) $ gives 
\begin{equation}
R=\frac{I_{s}}{E_{m}}.
\end{equation}
It is remarkable that we obtain a perfect subtraction of the ZPF, a result
in agreement with the quantum mechanical prediction. The result may be
generalized to the case where the signal intensity is not a constant, but a
known function of time. It would be enough to substitute $%
\int_{0}^{t}I_{s}(t^{\prime })dt^{\prime }$ for $I_{s}t$ in the above
equations, although the integrals would be more involved. More difficult
would be to treat the common case where the signal itself fluctuates (with a
correlation time of the order of the inverse of the frequency bandwidth). We
shall study that problem elsewhere.

We may now analyze coincidence counts in two detectors when the incoming
beams, with intensities $I_{1}(t)$ and $I_{2}(t)$ \textit{above the ZPF, }%
are correlated. The calculation is not difficult if the correlation time of
the signal is of the order of the typical time interval between detection
events, or larger. In these conditions we may assume that eq.$\left( \ref{R}%
\right) $ is still valid for each detector and the coincidence rate, with a
time delay $\tau ,$ will be 
\begin{equation}
R_{12}\propto \left\langle I_{1}(t)I_{2}(t+\tau )\right\rangle ,
\end{equation}
again in agreement with the quantum prediction. However the current
situation may not be that (see next section.) In practice the
crosscorrelation time of the signals is much shorter than the inverse of the
detection rate. Again the calculation in these conditions will be rather
involved and shall not be considered here.

\section{Discussion}

Our analysis shows that quantum vacuum fluctuations of the electromagnetic
field (or ZPF) may be efficiently subtracted by a\textit{\ }model which
assumes that the radiation is a classical (Maxwell) field including a
fluctuating ZPF, provided that the fluctuations of the signal have a large
enough correlation time in comparison with the correlation time of the ZPF.
This is usually the case in astronomical observations. In contrast, in
standard quantum optical experiments the fluctuations of the signal may have
a rather short correlation time. If the correlation time of the signal does
not fulfil the assumptions of the previous section, the presence of the ZPF
will probably give rise to deparatures from the standard quantum predictions
eqs.$\left( \ref{R}\right) $ and $\left( \ref{coin}\right) $, that is they
will produce some nonidealities in the behaviour of optical photon counters.
This is specially important when it is necessary to measure coincidence
counting rates with short time windows, as is frequent in quantum optical
experiments (e.g. optical tests of Bell\'{}s inequality). If this is the
case, our approach may provide an explanation for the difficulties of
performing loophole-free tests of Bell\'{}s inequality using optical
photons. As is well known all performed experiments suffer from the
''detection loophole'' \cite{laloe} and I conjecture that the cause might be
the existence of fundamental nonidealities in the behaviour of photon
counters.

I emphasize that, although our model is semiclassical, probably the main
properties of the model would be reproduced by a more rigorous quantum
treatment. Furthermore the difficulties for reaching an intuitive picture of
how detectors subtract the ZPF probably do not derive from quantum theory
itself, but from the use of approximations like first-order perturbation
theory or taking the limit of time t$\longrightarrow \infty $ in calculating
the probability of photon absorption per unit time. Indeed I have
conjectured elsewhere that excesive idealizations might be at the origin of
the difficulties for undersanding intuitively the paradoxical aspects of
quantum physics \cite{es}. Although simplifications are extremely useful for
calculations, they tend to obscure the physics.

\textbf{Acknowledgement. }I acknowledge financial support from DGICYT,
Project No. PB-98-0191 (Spain). I thank Trevor W. Marshall for pointing out
that the solution of eq.$\left( \ref{ode}\right) $ was wrong in a previous
version of the article.


\begin{thebibliography}{9}
\bibitem{lamb}  Prof. Lamb has expressed the opinion that his experiment
provided an empirical proof that the vacuum is not empty. This is related to
his statement that the photon is the quantum of the electromagnetic field,
but not a particle, see W. E. Lamb Jr., \textit{Appl. Phys.} \textbf{B 60},
77 (1995). The point is important for the present paper because it stresses
the difference between photon counting and detection of true particles, like
electrons or atoms. Indeed it is not difficult to detect atoms with
practically 100\% efficiency and no noise, but this is not the case with
photons.
\end{thebibliography}
\end{document}